\magnification=\magstep1
\hfuzz=40pt

\centerline{\bf The Hilbert Action in Regge Calculus}
\bigskip
\medskip
\centerline{Warner A. Miller}
\centerline{\it Theoretical Astrophysics Group (T-6, MS B288)}
\centerline{\it Theoretical Division}
\centerline{\it Los Alamos National Laboratory}
\centerline{\it Los Alamos, NM 87545}
\medskip
\centerline{\it 13 January 1997}
\bigskip
\bigskip
\centerline{\bf Abstract}
\medskip

The Hilbert action is derived for a simplicial geometry.  I recover
the usual Regge calculus action by way of a decomposition of the
simplicial geometry into 4-dimensional cells defined by the simplicial
(Delaunay) lattice as well as its dual (Voronoi) lattice.  Within the
simplicial geometry, the Riemann scalar curvature, the proper
4-volume, and hence, the Regge action is shown to be exact, in the
sense that the definition of the action does not require one to introduce
an averaging procedure, or a sequence of continuum metrics which were
common in all previous derivations.  It appears that the unity of
these two dual lattice geometries is a salient feature of Regge
calculus.

\bigskip
\line{\bf I. Gravity as Simplicial Geometry.\hfil}
\medskip

Einstein gave us a description of gravitation as geometry.  A
spacetime geometry can be represented arbitrarily closely by a lattice
of flat 4-dimensional triangles (4-simplices).  Such a simplicial
geometry is a straight-forward generalization to four dimensions of
the familiar architectural geodesic dome.  Here however, each building
block is not a flat triangle but a flat 4-simplex, and its intrinsic
geometry is Lorentzian rather than Euclidean.  It was Tullio Regge who
showed us, for the first time, how to encode the principles of General
Relativity onto a simplicial geometry --- a mathematics popularly
known today as Regge calculus.${}^{1-3}$

In his original paper, Regge introduced the notion of curvature on a
simplicial geometry and gave us the simplicial analogue of (1) the
Hilbert action, (2) Einstein equations, and (3) the Bianchi
identities.${}^1$ The objective of this paper will be to reinvestigate
the Regge calculus form of the Hilbert action and to provide a simple
geometric derivation of it.  The derivation presented here will shed
new light on the foundations of Regge calculus.

Although I am primarily interested in a 4-dimensional spacetime
realization of Regge calculus, the derivation of the action presented
here is valid for an arbitrary $\scriptstyle D$-dimensional simplicial
lattice.  In accord with this dimensional generality, I will derive
the action here in $\scriptstyle D$ dimensions.  Throughout this paper
I will refer to the simplicial lattice as a Delaunay lattice and to
its circumcentric dual lattice as a Voronoi lattice.  I have
introduced these lattices and terminology for two reasons.  First,
the Voronoi lattice appears to be a fundamental feature of Regge
calculus.  Second, by adopting this standard terminology Regge
calculus might become more accessible to a broader community and might
aid in cross fertilization of different scientific fields.  Precise
definitions of the Voronoi and Delaunay lattices can be found in the
literature.${}^4$ Finally, throughout this paper I use geometric units
where $G=c=1$.

There are currently two derivations of the Hilbert action in Regge
calculus.  The first was introduced by Regge and involved an averaging
procedure.${}^1$ The second was by R.~Friedberg and T.~D.~Lee which
employed an embedding of the simplicial lattice into a higher
dimensional Euclidean geometry and a sequence of continuum metrics
which converged to the simplicial spacetime.${}^5$ While both
derivations are consistent, and the latter provides a useful
mathematical linkage between Regge calculus and the continuum, they
lead one to believe that the Regge-Hilbert action is an approximate,
truncated expression.  I will demonstrate in this paper that the
notions of parallel transport, curvature and proper volumes are
precisely defined in Regge calculus; and this leads to an internally
exact geometric construction of the Regge-Hilbert action.
Furthermore, the emergence of the Voronoi lattice as a natural facet
of Regge calculus is (1) consistent with an earlier and independent
derivation of the simplicial form of the Einstein equations (Regge
equations) as a sum of moment-of-rotation trivectors,${}^6$ and is (2)
in conformity with a previous derivation of the $R^2$ action terms in
Regge calculus.${}^7$ What I wish to emphasize in this paper is that
the principles of general relativity can be applied directly to the
simplicial spacetime geometry to yield an exact expression for the
Regge calculus form of the Hilbert action.

\bigskip
\line{\bf II.  Dual Voronoi--Delaunay Derivation of the Hilbert Action 
               in Regge Calculus.\hfil}
\medskip

In a $\scriptstyle D$-dimensional Delaunay lattice geometry the curvature is
concentrated on the $\scriptstyle (D-2)$-dimensional simplices.  We
refer to these co-dimension two simplices as hinges.  In order to
transcribe the Hilbert action into Regge calculus, we need to define
the scalar curvature tensor, $^{\scriptscriptstyle (D)}\!R$, and the
$\scriptstyle D$-dimensional volume elements,
$d\,{}^{\scriptscriptstyle(D)}\!V_{\rm proper}$.

$$
I_{\rm Hilbert} = {1\over 16\pi} \int {}^{\scriptscriptstyle (D)}\!R\; 
                                     d\,{}^{\scriptscriptstyle(D)}\!V_{\rm proper}.
\eqno(1)
$$

Curvature can be revealed in a number of ways ranging from geodesic
deviation to surface area deficit.  I know of no more appropriate
vehicle toward the understanding of curvature on a simplicial geometry
than the parallel transportation of a vector around a closed loop.
$$
K = \pmatrix{\hbox{Gaussian}\cr
             \hbox{Curvature}\cr} = 
{ \hbox{Angle that Vector is Rotated}\over 
  \hbox{Area Circumnavigated} }
\eqno(2)
$$
However, in our $\scriptstyle D$-dimensional simplicial geometry the
curvature at each hinge takes the form of a conical singularity.
Consequently, the deflection of a gyroscope when carried one complete
circuit around a closed loop encircling a given hinge ($h$) is
independent of the area of the loop.  This circumstance presents us
with our first dilemma.  How do we define the curvature of a hinge?
Equivalently, we may ask, what area, content or measure do we assign
to a given hinge, $h$?  A natural measure for $h$ emerges if we
consider the collection of points in the Delaunay lattice geometry
closer to $h$ than to any other hinge.  If there are $N_h$
$\scriptstyle D$-dimensional simplices sharing hinge $h$, then the set
of points thus assignable to $h$ naturally form $N_h$ simplices.  These
simplices are formed by taking the Voronoi polygon $h^*$ dual to hinge
$h$ and connecting its $N_h$ vertices to the $\scriptstyle (D-1)$
vertices of $h$.  Each corner on $h^*$ lies at the circumcenter of one
of the $\scriptstyle D$-simplices in the Delaunay geometry which share
hinge $h$.  By construction, the Voronoi polygon $h^*$ is orthogonal
to hinge $h$.  We use this orthogonality (Appendix) to show that the
$D$-volume of these $N_h$ simplices, determined by both the Delaunay
lattice (hinge, $h$) and its dual Voronoi lattice (polygon, $h^*$), is
proportional to the simple product of (1) $A_h$, the $\scriptstyle
(D-2)$-dimensional volume of hinge $h$, and (2) $A^*_h$, the area of
the Voronoi polygon $h^*$.
$$
\Delta\,{}^{\scriptscriptstyle (D)}\!V = 
{\displaystyle { 2\over D(D-1)}}\  A_h A^*_h.
\eqno(3)
$$
The proportionality constant is related to the dimension,
$\scriptstyle D$, of our lattice geometry.

These dual Voronoi-Delaunay volume elements provide us with a natural 
decomposition for our Hilbert action (Eq.~1) as well as a natural structure 
to define the scalar curvature of any hinge, $h$.
$$
\int d\,{}^{\scriptscriptstyle(D)}\!V_{\rm proper}
\Longrightarrow \sum_{hinges,\atop h}
{\displaystyle {2\over D(D-1)}}\  A_h A^*_h.
\eqno(4)
$$

To define the curvature concentrated at hinge $h$ it seems natural to
assign the area $A^*_h$ to the hinge $h$.  If we parallel transport a
vector around the perimeter of $h^*$ it will have traversed the flat
geometry of the interior of each of the $N_h$ simplices sharing $h$.
Furthermore, it will return rotated in a plane perpendicular to $h$
(hence parallel to $h^*$) by an angle $\varepsilon_h$.  This deficit
angle $\varepsilon_h$ is defined by the hyperdihedral angles ($\theta_i$) 
of each of the $N_h$ simplices of the Delaunay lattice sharing hinge 
$h$.
$$
\varepsilon_h = 2\pi-\sum_{i=1}^{N_h} \theta_i
\eqno(5)
$$
In particular, $\theta_i$ is the angle between the $(D-1)$-dimensional
faces of simplex $i$ hinging on $h$.  The Gaussian curvature for $h$
is thus,
$$
K_h = {\varepsilon_h\over A^*_h}.
\eqno(6)
$$
Since the plane of rotation and the rotation bivector are both
orthogonal to hinge $h$, and since there are no other curvature
sources associated with $h^*$ then the Riemann scalar curvature is
simply proportional to $K_h$.  
$$
\pmatrix{\hbox{Riemann Scalar Curvature}\cr
         \hbox{Associated with Hinge,}\  h} =
{}^{\scriptscriptstyle (D)}\!R_h = 
{D(D-1)\varepsilon_h\over A^*_h}.
\eqno(7)
$$

Therefore, one can translate the Hilbert action into its Regge
calculus form by using the dual Voronoi-Delaunay construction
presented here through equations (4) and (7).
$$
I_{\rm Hilbert} = {1\over 16\pi} \int {}^{\scriptscriptstyle (D)}\!R\; 
                  d\,{}^{\scriptscriptstyle(D)}\!V_{\rm proper}
\Longrightarrow 
{1\over 16\pi} \sum_{{\rm hinges,}\atop h} 
\underbrace{
\left({D(D-1)\varepsilon_h\over A^*_h}\right)}_
{\displaystyle {}^{\scriptscriptstyle (D)}\!R}
\underbrace{\left({\displaystyle {2\over D(D-1)}}\ A_h A^*_h\right)}_
{\displaystyle \Delta\,{}^{\scriptscriptstyle (D)}\!V}
\eqno(8)
$$
We see that the Voronoi area $A^*_h$ cancels as well as the
dimensionality constants $D(D-1)$.  We are left with the usual Regge
calculus expression for the Hilbert action.
$$
I_{\rm Regge} = {1\over 8\pi}  \sum_{{\rm hinges,}\atop h} 
A_h \varepsilon_h.
\eqno(9)
$$

\bigskip
\line{\bf III. Voronoi Lattice as Equally as Important as the\hfil} 
\line{\bf \ \ \ \ \  Simplicial Delaunay Lattice in Regge Calculus.\hfil}
\medskip

Two new insights into the inner workings of Regge calculus are
obtained in this paper.  First, we showed that the principles of
General Relativity can be applied directly to the lattice geometry to
yield an internally exact expression for the Riemann curvature, proper
volume; and hence, the Hilbert action.  Second, we showed that Regge
calculus required a unity of the simplicial geometry (Delaunay
lattice) and its dual geometry (Voronoi lattice).  The complex of
simplices we formed by connecting a given hinge in the Delaunay
lattice to its corresponding Voronoi polygon appears to be a natural
geometric structure to support the Hilbert action.  The orthogonality
between these two lattices provided a compact expression for the
volume of this complex of simplices which then led to an interesting
cancellation of terms in the action.  In the final expression for the
action (Eq.~9) we see nowhere any trace of the Voronoi lattice ---
the areas of the Voronoi polygons all cancelled.  This apparently deep
union of these two lattices can be observed in an action--independent
derivation of the simplicial version of the Einstein equations.${}^6$
In particular, the author needed, in this earlier work, to introduce
the Voronoi lattice geometry in order to construct the Cartan--derived
Einstein equations as well as the corresponding Bianchi identities.

Regge calculus provides us with a geometric, discrete and finite
rendering of one of the most beautiful theories in physics ---
Einstein's geometric theory of gravitation.  One wonders what facets
of this theory one can distill out to gain an even deeper
understanding of nature?  Perhaps the geometric duality highlighted in
this article is one such indicator?

\bigskip
\line{\bf IV. Acknowledgements \hfil} 
\medskip

I wish to thank Harold Trease for valuable discussions regarding
Voronoi lattices, and especially John A. Wheeler for his scientific 
support and continued encouragment to push forward the frontiers of
Regge calculus.  This work was supported, in part, under an LDRD/GSSA
grant from Los Almaos National Laboratory.

\bigskip
\line{\bf Appendix: Definition of $A^*_h$ and Proof of Eq.~3.\hfil}
\medskip

To derive Eq.~3, we consider a $\scriptstyle (D-2)$--dimensional hinge, $h$, in a 
$\scriptstyle D$--dimensional simplicial geometry.  
$$
h=\{H_0,H_1, \ldots, H_{(D-2)}\}.
\eqno(A1)
$$
Here, the $H$'s label the vertices of the hinge.  We assume that there
are $N_h$ $\scriptstyle D$-simplices sharing hinge $h$. This
collection of $N_h$ simplices $\{{\cal S}_0,{\cal S}_1, \ldots, {\cal
S}_{N_h-1}\}$ is formed by connecting the $H$'s to a chain of $N_h$
vertices, $\{S_0, S_1, \ldots, S_{N_h-1}\}$, where
$$
{\cal S}_j = \{H_0,H_1,\ldots,H_{(D-2)},S_j,S_k\}, 
\qquad\hbox{and}\quad k=j+1\bmod N_h. 
\eqno(A2)
$$
Let $O_h$ be the circumcenter of hinge $h$. In other words, $O_h$ is
the point in the ${\scriptstyle (D-2)}$--dimensional plane formed by
$h$ that is equidistant from all the $H$'s.  Similarily, let $C_j$ be
the circumcenter of the corresponding simplex ${\cal S}_j$, for each
$j$.  Then the Voronoi polygon $h^*$ dual to hinge $h$ is simply the
the collection of these $N_h$ circumcenters,
$$
h^* = \{C_0,C_1,\ldots,C_{N_h-1}\}.
\eqno(A3)
$$
The area of $h^*$, which we have referred to as $A^*_h$ is formed by
connecting $O_h$ to the vertices of $h^*$. This breaks the Voronoi
polygon ($h^*$) into $N_h$ triangles, $h^*=\{h^*_i, \quad i=0,1,\ldots,
N_h-1\}$, where $h^*_i=\{O_h,C_i,C_j\}$ with $j=i+1\bmod N_h$. Here 
$$
A^*_h = \sum_{i=0}^{N_h-1} A^*_{h_i},
\eqno(A4)
$$
where $A^*_{h_i}$ is the area of triangle $h^*_i$. Each of these triangles 
$A^*_{h_i}$ is orthogonal to $h$.

When we connect hinge $h$ with the Voronoi polygon $h^*$ we form a
complex of $N_h$ simplices, $\{{\cal V}_i,\quad i=0,1,\ldots,N_h-1\}$.
Each of these simplices (${\cal V}_i$) is formed by connecting the
$i$'th edge of $h^*$ to the hinge $h$,
$$
{{\cal V}_i} = \{H_0,H_1,\ldots,H_{(D-2)},C_i,C_{i+1\bmod N_h}\}.
\eqno(A5)
$$
To derive Eq.~3 we need only show that the D-volume of simplex ${\cal
V}_i$ is given by,
$$
V_i = {\displaystyle {2\over D(D-1)}}\ A_h A^*_{h_i}.
\eqno(A6)
$$

When we insert the circumcenter ($O_h$) of hinge $h$ into simplex
${\cal V}_i$ and connect it to its $D+1$ vertices, we form $D-1$ new
D-simplices.  Now,
$$ 
V_i = \sum_{j=0}^{D-2} V_{i_j},
\eqno(A7)
$$
and ${\cal V}_{i_j}$ is formed by replacing the vertex $H_j$ of simplex 
${\cal V}_i$ with the circumcenter ($O_h$) of the hinge $h$.  Now we are in
a position to verify Eq.~A6 and thus Eq.~3.

Using the volume formula for a $D$--dimensional simplex we notice that,
$$
V_{i_j} = {1\over D!} \left( \overrightarrow{O_hH_0}     \wedge
                            \overrightarrow{O_hH_1}     \wedge \ldots
                            \overrightarrow{O_hH_{j-1}} \wedge
                            \overrightarrow{O_hH_{j+1}} \wedge \ldots
                            \overrightarrow{O_hH_{D-2}} \wedge
                            \overrightarrow{O_hC_i}     \wedge
                            \overrightarrow{O_hC_{i+1\bmod N_h}}
                     \right),
\eqno(A8)
$$
for all $i=0,1,\ldots,N_h-1$.  However, triangle $h^*_i$ is orthogonal
to hinge $h$.  Therefore we can factor the last wedge product out of
Eq.~A8.
$$
V_{i_j} = {1\over D!} \underbrace{
                     \left( \overrightarrow{O_hH_0}     \wedge
                            \overrightarrow{O_hH_1}     \wedge \ldots
                            \overrightarrow{O_hH_{j-1}} \wedge
                            \overrightarrow{O_hH_{j+1}} \wedge \ldots
                            \overrightarrow{O_hH_{D-2}} \right) 
                                }_{\displaystyle (D-2)! A_{h_j}}
                     \underbrace{
                     \left( \overrightarrow{O_hC_i}     \wedge
                            \overrightarrow{O_hC_{i+1\bmod N_h}}
                     \right)
                                }_{\displaystyle 2A^*_i}.
\eqno(A9)
$$
However, this last underbraced term is just twice the area of triangle
$h^*_i$, and the first term is just $(D-2)!$ times that fraction of
the $(D-2)$-volume ($A_{h_j}$) of the hinge $h$ formed by replacing
vertex $H_j$ with its circumcenter $O_h$. Therefore,
$$
V_{i_j} = {2\over D(D-1)} A_{h_j} A^*_i,
\eqno(A10)
$$
and by using this equation and Eq.~A7 we have verified Eq.~A6. This
together with Eq.~A4 completes the derivation of Eq.~3 and the definition
of the area of $h^*$.

\bigskip
\line{\bf References.\hfil}
\medskip
\item{${}^1$} T. Regge, ``General Relativity Without Coordinates,'' {\it Nuovo
Cimento} {\bf 19} (1961) 558-571.
\medskip
\item{${}^2$} C. W. Misner, K.S. Thorne and J. A. Wheeler, {\it Gravitation} 
(Freeman, New York, 1973), Ch.~42.
\medskip
\item{${}^3$} R. M. Williams and P. A. Tuckey, ``Regge Calculus: A Brief Review and
Bibliography,'' {\it Class. Quantum Grav.} {\bf 9} (1992) 1409-1422.  
\medskip
\item{${}^4$} A. Okabe, B. Boots and K. Sugihara, {\it Spatial Tesselations,
Concepts and Applications of Voronoi Diagrams} (John Wiley \& Sons, New York, 1992),
Ch.~2.
\medskip
\item{${}^5$} R. Friedberg and T. D. Lee, ``Derivation of Regge's Action from 
Einstein's Theory of General Relativity,'' {\it Nucl. Phys.} {\bf B242}
(1984) 145-166.
\medskip
\item{${}^6$} W. A. Miller, ``The Geometrodynamic Content of the Regge 
Equations as Illuminated by the Boundary of a Boundary Principle,'' {\it
Found. Phys.}  {\bf 16} (1986) 143-169.
\medskip
\item{${}^7$} H. W. Hamber and R. M. Williams, ``Higher Derivative
Quantum Gravity on a Simplicial Lattice,'' {\it Nucl. Phys.} {\bf B248}
(1984) 392-414.

\bye